\documentstyle[preprint,eqsecnum,aps]{revtex}
\begin{document}
\draft
\preprint{SU-GP-93/5-4, gr-qc/9307003}
\title{Models of Particle Detection in Regions of Spacetime}
\author{Donald M. Marolf\cite{Marolf}}
\address{Physics Department, Syracuse University,
Syracuse, New York 13244} \date{July 6, 1993}
\maketitle

\begin{abstract}
We investigate two models of measuring devices
designed to detect a non-relativistic free particle in a
given region of spacetime.  These models predict different
probabilities for a free quantum particle
to enter a spacetime region $R$ so that this notion
is device dependent.  The first model is of a von Neumann coupling
which we present as a contrast to the second model.  The second model
is shown to be related to probabilities
defined through partitions of configuration space
paths in a path integral.  This study thus provides insight into the
physical situations to which such definitions of probabilities are
appropriate.

\end{abstract} \pacs{3.65.Bz}

\section{Introduction}
\label{Int}

The theory of measurement in quantum mechanics has been a source of
discussion and controversy for as long as the subject has existed.
Indeed, the Syracuse University Library Catalogue system returns
a list of 13 books published since 1968 in response to the keyword
inquiry ``quantum measurement."  Such a list may at best be
considered abbreviated as it neglects works predating 1968, such as
von Neumann's {\it Mathematical Foundations of Quantum Mechanics}
\cite{Neumann}, and numerous
research papers not found in books.
Within this immense volume of material, a number
of important contributions have been made by investigating
specific examples (or at least models) of experimental situations;
in other words through a study of measuring devices.  These
contributions include
the classic works of Mott \cite{Mott}, von Neumann \cite{Neumann},
Bohr and Rosenfeld \cite{BR}, and DeWitt \cite{Bryce} as well as
the more recent studies of detectors in quantum field theories
which are summarized by N. Birrell and P.C.W. Davies in
{\it Quantum Fields in Curved Space} \cite{BD}.

The basic procedures followed in these studies is to first
describe an appropriate classical measuring device
and to then quantize the coupled system
consisting of both the apparatus and the system
to be measured.  In the above examples, the
couplings considered single out operators associated with the
measured system that can be said to be
``measured" so that the ``experiment" may then be given an
interpretation
by the textbook procedure in which the total state of the
system is decomposed as a sum of eigenvectors of this operator
and the ``probabilities" to measure various eigenvalues are given
by the coefficients in this sum.

At this point a comment is in order concerning the use of
quotation marks in this paper.
It is the author's intention to use such punctuation to avoid a
discussion of the interpretation of quantum mechanics
as the precise meaning of the words ``probabilities,"
``measurement," and even ``outcome" for each reader will be
influenced by the interpretation to which he/she subscribes.
Rather than single out one set of such definitions for use
here or use complicated notation to distinguish between the
various meanings present in the literature at large, we
choose to use these words
for any of the possible meanings but to recognize that an
ambiguity exists by enclosing the words in quotation marks
for {\it every} such use.
When such a term appears in the text,
the intended meaning should be clear from the context (as in the
above paragraph) and is ultimately determined by the mathematical
steps used to calculate the corresponding ``probabilities".

Suppose then that we wish to predict whether a quantum free
particle will be found in some region $R$ of spacetime --
a case that has received some
attention \cite{Hartle,first,YT}.  An analysis in the spirit of
\cite{Mott,Neumann,BR,Bryce,BD}
would begin with the description of a classical detector
and then proceed with a quantization of the coupled
detector/particle system. A study of the quantum detector
after the experiment is completed determines
the ``probabilities" of the outcomes through the norms
of the relative states \cite{Ev}\footnote{We note that,
while the terminology of Everett \cite{Ev} is
convenient here, the use of such terminology does not
imply that the following analysis assumes the Everett
interpretation of quantum mechanics.}  associated with possible
responses of the
device and it may be that the results are highly dependent on the
device.

We therefore study models of two such detectors.
The first such model (A) is
presented in section \ref{I} and leads to a definition of
probability of the
familiar type given by projections of the free particle state
onto parts of the spectrum of some Hermitian operator.
On the other hand, the model (B)
presented in section \ref{II} is related to a definition of
such probabilities which has been previously proposed
\cite{Hartle,first,YT} in terms of partitions of paths
in a path integral.  We note that
these two definitions are not equivalent, so that this study
may be used to develop an understanding of the physical
situations to which such definitions of probabilities
are and are not appropriate.

This discussion will constitute the bulk of the text,
after which we will
close with a discussion of the main points and two appendices.
Appendix A
presents a brief discussion of bubble chambers in order
to contrast this device with models
A and B.  Appendix B reviews the ``ideal measurement limit"
which we use in our models.

\section{Single Measurement Model}
\label{I}

For definiteness, we would like to analyze the detection of a free
non-relativistic particle in 1+1 dimensions in the spacetime region
$R$ to the right of $x=0$ between the times $T_1$ and
$T_2$.  Our detector will be defined by a pointer which moves
to the right when the particle is present in
region $R$.   We
will then say that the particle has or has not entered
$R$ based on whether or not the pointer has moved between times
$T_1$ and $T_2$.

Specifically, we will assume that the coupling is
of a von Neumann type associated with measurement of $\chi =
\int_{T_1}^{T_2} dt \ \theta(x)$ during the time interval
$(T_1,T_2)$ so that if $A_g(t)$ is the pointer value at time $t$
when coupled with
strength $g$ to the particle and
under the retarded boundary conditions
$A_g(T_1) = A_0(T_1)$ we have $A_g(T_2) = A_0(T_2) + g\chi_0$.
Similarly, $\chi_0$ is the undisturbed
($g$=0) value of $\chi$\footnote{It is interesting to note that it
difficult to construct such couplings through an action or
Hamiltonian principle without first expressing $\chi_0$ in terms
of the operators $P(T_1)$ and $X(T_1)$ through the uncoupled
equations of motion.
That is, without essentially
reducing it to an operator defined at the single time $T_1$.}.
Note that $\chi$ is just
the total time spent by the particle in
region $R$ so that this is consistent with our decision to say
that the particle entered region $R$ if and only if the pointer
moves during the experiment.

We will also assume that the
free ($g$=0) pointer evolution is trivial: $A_0(t) = A_0(t')$
and $\pi_0(t) =
\pi_0(t')$ where $\pi$ is the quantity canonically conjugate to $A$.
This may be
regarded as either an unphysical but useful mathematical
model or as the limit of a pointer that is extremely massive
compared to its momentum, {\it i.e.} such that
$P^2(T_1-T_2)/M\hbar << 1$.  Alternatively, we might take
the free pointer Hamiltonian to be $A^2$
($A$ might be the momentum of some other free particle) in
which case $A$ would still be conserved when the device
and free particle do not interact.  We require conservation of
$A$ for the
uncoupled device since a pointer set initially to zero should
remain at zero if no particle is present.

The quantization of this model is the straightforward.
As we will want to compare the disturbed (finite $g$) and
undisturbed ($g=0$) cases, we will in fact construct a
quantum theory for each $g$.
We will use the Heisenberg picture and choose representations
carried by Hilbert spaces isomorphic to each other and
to ${\cal L}^2({\cal R}^2) = {\cal L}^2({\cal R};x)
\otimes {\cal L}^2({\cal R};A)$.  Of particular use will be the
isomorphism $I^-_g$ from the Hilbert space ${\cal H}_0$ of the
uncoupled system
to the Hilbert space ${\cal H}_g$ of the system with coupling
strength $g$ that satisfies
\begin{equation}
\label{iso}
(I^-_g)^{-1} \hat{\cal O}_g(t) I^-_g = \hat{\cal O}_0(t)
\end{equation}
for any operator $\hat{\cal O}_0(t)$ built from the basic operators
$\hat{X}_0(t)$, $\hat{P}_0(t)$,
$\hat{A}_0(t)$, and $\hat{\pi}_0(t)$ (the particle
and pointer positions and momenta at the time $t$ in the uncoupled
system) and the operator $\hat{\cal O}_g(t)$ built in exactly the
same way but from the basic operators
$\hat{X}_g(t)$, $\hat{P}_g(t)$, $\hat{A}_g(t)$, and $\hat{\pi}_g(t)$
at some time $t \leq T_1$.  Because the coupling begins only at
time $T_1$, the same isomorphism satisfies \ref{iso}
for all $t \leq T_1$ and this isomorphism corresponds to
our classical use of retarded boundary conditions (a fixed
$g$-independent initial condition) in comparing the coupled and
uncoupled systems.

Because our coupling is to be of the von Neumann type
associated with
measurements of $\chi$, we take the quantum dynamics to be defined
so that
\begin{equation}
\label{dynA}
\hat{\cal O}_g(T_2) = I_g^- \exp(-ig\hat{\pi}_0 \hat{\chi}_0)
\ \hat{\cal O}_0
(T_2)\  \exp(ig\hat{\pi}_0 \hat{\chi}_0) (I_g^-)^{-1}
\end{equation}
Here, $\hat{\chi}_0$ is the operator $\int_{T_1}^{T_2} dt\
\hat{\theta}_0(x(t))$
in ${\cal H}_0$ and $\hat{\theta}_0(x(t))$ is the
projection onto the positive spectrum of
$\hat{X}_0(t)$.  Note that $ I^-_g \widehat{\chi}_0
(I^-_g)^{-1} \neq \hat{\chi}_g$ since $\chi$ is built from the
basic operators between times $T_1$ and $T_2$.

As is  consistent with our use of retarded boundary conditions, we
will assume that the state of our system contains
no correlations between the particle and
detector at time $T_1$.  That is, we take $|\psi\rangle$
to be of the form
\begin{equation}
|\psi\rangle = |\phi_p\rangle \otimes |\phi_D\rangle
\end{equation}
in terms of the factorization ${\cal H}_g = {\cal H}_{g,p:T_1}
\otimes {\cal H}_{g,D:T_1}$ of the total Hilbert space ${\cal H}_g$
into a Hilbert space ${\cal H}_{g,p:T_1}$
associated with the particle at time $T_1$ and a Hilbert space
${\cal H}_{g,D:T_1}$ associated with the detector at time $T_1$.
Here $|\phi_p\rangle$ and $|\phi_D\rangle$ are normalized states in
${\cal H}_{g,p:T_1}$ and ${\cal H}_{g,D:T_1}$.

We now note that since $\hat{\chi}_0$ is defined for the uncoupled
system (in which the factorization ${\cal H}_0 = {\cal H}_{0,p:t}
\otimes {\cal H}_{0,D:t}$ is $t$-independent),
it can be written as the direct product:
\begin{equation}
\label{chiform}
\hat{\chi}_0 = \hat{\chi}_f \otimes \openone_{0,D}
\end{equation}
for some $\hat{\chi}_f: {\cal H}_{0,p:T_1}\rightarrow
{\cal H}_{0,p:T_1}$ where $\openone_{0,D}$ is
the identity operator in ${\cal H}_{0,D:T_1}$.
Similarly, for any operator ${\cal O}_0$ of the form \ref{chiform},
it will be convenient to define a corresponding operator
${\cal O}_f$ by ${\cal O}_0 = {\cal O}_f \otimes \openone_{0,D}$.

The evolution \ref{dynA} then leads in the usual way
(see \cite{Neumann,meas}) to the statement
that in the ideal measurement limit\footnote{A
brief review of this limit is
presented in appendix \ref{limitA} for comparison with our
discussion of model B in section \ref{limitB}.} we find
\begin{equation}
\label{distributions}
\langle \psi |
\hat{\Pi}_{A(T_2)=A'} dA' |\psi\rangle \rightarrow
|\langle \chi_f=A'| I^-_{g,f}|\phi_p\rangle|^2
d\mu_{\hat{\chi}_f}(A')
\end{equation}
of ``probability densities",
where $|\chi_f=A'\rangle \langle \chi_f=A'|
d\mu_{\hat{\chi}_f}(A')$ and $\hat{\Pi}_{A(T_2) = A'}dA'$
are the spectral measures evaluated at $A'$
of the operators $\hat{\chi}_f$ and $\hat{A}(T_2)$ respectively
and $I^-_{g,f}$ is the isomorphism from ${\cal H}_{g,p:T_1}$
to ${\cal H}_{0,p:T_1}$ induced by the isomorphism
$I^-_g: {\cal H}_0
\rightarrow {\cal H}_g$.   Equivalently, the decoherence functional
$\langle \psi |
\hat{\Pi}_{A=A'}dA'\  \hat{\Pi}_{A=A''}dA'' |\psi\rangle$
converges in this limit to
\begin{equation}
\label{naive}
{\cal D}(A',A'') =
\langle \phi_p |
\hat{\Pi}_{\chi_f=A'}
d\mu_{\hat{\chi}_f}(A')\  \hat{\Pi}_{\chi_f=A''}
d\mu_{\hat{\chi}_f}(A'') |\phi_p\rangle
\end{equation}
A general review of decoherence
functionals is given
in \cite{Hartle,Griffiths,GM} and in other works.
The results are equivalent and
decoherence is trivial because each decoherence functional
involves only commuting projection operators.

As defined by
the distribution \ref{distributions}, the ``probability" of
finding $A=0$ with arbitrarily high
accuracy vanishes unless the spectral distribution of
$\hat{\chi}_f$ is singular.  That is, it
vanishes unless $\hat{\chi}_f$
has a normalizable eigenstate in ${\cal H}_{0,p:T_1}$
with eigenvalue zero.  That this is
{\em not} the case can be seen by taking the expectation value of
$\hat{\chi}_f$ in any state $|\phi\rangle \in {\cal H}_{0,p:T_1}$:
\begin{eqnarray}
\langle \phi|\hat{\chi}_f|\phi \rangle &=&
\langle \phi| \int_{T_1}^{T_2} dt \ \hat{\theta}_f (x(t))|\phi
\rangle  = \int_{T_1}^{T_2} dt \ \langle \phi|\hat{\theta}_f
(x(t))|\phi \rangle \cr &=& \int_{T_1}^{T_2} dt
\int_0^{\infty} dx \phi^*(x;t) \phi(x;t)
\end{eqnarray}
where $\phi(x;t) = \langle \phi | x;t \rangle$ where
$|x;t\rangle$ is the eigenvector of $\hat{X_f}(t)$ in
${\cal H}_{0,p:T_1}$ with
eigenvalue $x$.  Since an arbitrary state
$|\phi \rangle$ can have $\phi(x;t)$ vanish for all $x \geq 0$
only for isolated times $t$, this integral is
always greater than zero so that $\hat{\chi}_f$ does not annihilate
any state in ${\cal H}_{0,p:T_1}$ and therefore does not have any
normalizable eigenvectors with eigenvalue zero.
It follows that in the limit of ideal measurement,
the particle is detected by device A with
``probability" one independent of the initial
state $|\phi_p\rangle$.

\section{A Multiple Measurement Model}
\label{II}

While we based device A on the von Neumann measurement of a single
classical quantity $\chi_0$, we will follow a different strategy
for device B and build it instead from {\em many} von
Neumann measurements. While this may seem like an unnatural
procedure from the viewpoint of measurement theory in quantum
mechanics (where we usually follow a ``one question, one
measurement" principle), this approach will lead to a definition
of the probability for a free particle to enter the region $R$
given by partitions of paths in
a path integral \cite{Hartle,first,YT}.
We therefore
pursue it primarily as a
way of understanding the type of physical situations to which such a
definition is appropriate.

Subsection A gives an overview of the classical model which is
quantized in subsection B.  Subsection B then shows that
the corresponding ``probabilities" can be related to a path
integral that sums only over paths which avoid the region $R$.
Subsection C relates model B to the discussion given in
\cite{Hartle,first,YT} and subsection D analyzes the difference
between our approach and that of \cite{Hartle,first,YT}
in light of the fact that device B disturbs the
system it measures.

\subsection{Model}

We define model B by coupling
${\cal P}_{\lambda,\tau}(t) =
\delta(x(t) - \lambda) \delta (t - \tau)$
to a pointer for each $(\lambda,\tau)$ in $R$.  Note
that these are ``explicitly time dependent operators" in
the sense that ${{d{\cal P}} \over {dt}} \neq \{ {\cal P},
H\}$ and that they are related to $\chi$ of section \ref{I} through
\begin{equation}
\chi = \int_{-\infty}^{\infty}dt\ \int_R d\lambda d\tau \
{\cal P}_{\lambda,\tau}
\end{equation}
It follows that $\chi$ is nonzero exactly when
\begin{equation}
\label{response}
\int_R d\lambda d\tau \ \phi(\lambda,\tau) {\cal P}_{\lambda,\tau}
\end{equation}
is nonzero for some smooth function $\phi$ and that
classically devices A and B detect
particles under identical conditions.

Thus, for each $(\lambda,\tau) \in R$, we introduce a pointer
coordinate $A_{\lambda,\tau}$ and a conjugate
momentum $\pi_{\lambda,\tau}$ and define model B through
the action:
\begin{equation}
\label{B}
S_{B} = S_{free} -
g\int_{-\infty}^{\infty} dt \int_{(\lambda,\tau) \in R}
d\lambda d\tau \ \pi_{\lambda,\tau} {\cal P}_{\lambda,\tau}
+ \int_{-\infty}^{\infty} dt
\int_{(\lambda,\tau) \in R}  d\lambda d\tau \ \pi_{\lambda,\tau}
\dot{A}_{\lambda,\tau}
\end{equation}
where again we have assumed that the pointers have
trivial free evolution.
We then interpret the result of this
``experiment" by saying that the
particle entered region $R$ if and only if some pointer
is disturbed during the
time interval.
Note that each coupling takes place instantaneously and,
as a result, each pointer $A_{\lambda,\tau}$
coupled to ${\cal P}_{\lambda,\tau}$ responds to the same value of
${\cal P}_{\lambda,\tau}$ as it would have if the $(\lambda,\tau)$
coupling were not present.  However, the $(\lambda,\tau)$ coupling
will effect the value of any ${\cal P}_{\lambda',\tau'}$ measured
afterwards since from \cite{Peierls} we see that the change
induced in $I(\lambda_2,\tau_2) = \int_{\infty}^{\infty} dt
\ {\cal P}_{\lambda_2,\tau_2}$
by the $(\lambda_1,\tau_2)$ coupling is given by
\begin{equation}
\label{effects}
D^-_{I(\lambda_1,\tau_1)}I(\lambda_1,\tau_1) = (I(\lambda_1,\tau_1),
I(\lambda_2,\tau_2)) \ \theta(\tau_1,\tau_2) \neq 0
\end{equation}
where the bracket in \ref{effects} is the Peierls
bracket\footnote{The Peierls
bracket (A,B)
may be thought of as the Poisson bracket extended to quantities
evaluated at different times by using the equations of motion.} and
we have used the notation of \cite{BryceLH}.  In language
that is more familiar from quantum mechanics,
this experiment attempts
to measure a set of operators that do not commute with
each other and
the result is that the measurement of one necessarily effects
the outcome of the others.  In this sense then, model B describes
a device which always disturbs the system with which it interacts.

\subsection{Quantization}
\label{limitB}

We again define the quantum system for each value of the coupling
constant $g$ in terms of a Hilbert space ${\cal H}_g$ isomorphic
to ${\cal L}^2(X) \otimes {\cal L}^2(A)$, although now
${\cal L}^2(A)$
is the Hilbert space ${\cal L}^2(\{A: R\rightarrow {\cal R} \})$
defined by some measure on the space of
real functions on $R$.  Since the coupling exists only for
$T_1 \leq t\leq T_2$, we have another ``retarded"
isomorphism $I_g^- : {\cal H}_0 \rightarrow
{\cal H}_g$ defined by
\begin{equation}
(I_g^-)^{-1} {\cal O}_g(t) I_g^- = {\cal O}_0(t)
\end{equation}
for all operators ${\cal O}_0(t)$ and ${\cal O}_g(t)$ such
that they are built in the same way from $\{X_0(t), P_0(t),
A_{\lambda,\tau,0}(t), \pi_{\lambda,\tau,0}(t)\}$ and
$\{X_g(t), P_g(t), A_{\lambda,\tau,g}(t),
\pi_{\lambda,\tau,g}(t)\}$ respectively.  The state of
the system will again be assumed to be uncorrelated at
time $T_1$: $|\psi\rangle = |\phi_p\rangle \otimes |\phi_D\rangle
\in {\cal H}_{g,p:T_1} \otimes {\cal H}_{g,D:T_1}$ for normalized
states $|\phi_p\rangle$ and $|\phi_D\rangle$.

However, as we
are not considering the reaction of a pointer to
a single undisturbed quantity, we cannot simply
use the standard results to draw conclusions about
``probabilities" (or ``decoherence") in the ideal
measurement limit but must derive the appropriate
results ourselves.  We will model this derivation
after the discussion in
appendix B and begin by assuming that the initial
state $|\phi_D\rangle$
of our device is characterized by a width $\sigma$ and is
given explicitly by:
\begin{equation}
|\phi_D\rangle  = {1 \over N} \int
\prod_{\lambda,\tau}dA_{\lambda,\tau}
\exp(-\int_R d\lambda d\tau \ A^2/2{\sigma}^2) |A;T_1\rangle
\end{equation}
in terms of the eigenstates $|A;T_1\rangle$ of
$A_{\lambda,\tau}(T_1)$
and where $N$ is chosen so that $\langle \phi_D|\phi_D\rangle =1$.

We would like to say (in some sense) that
detection of a particle in the region $R$
at a confidence level $n$ occurs when the
response of the device is
greater than $n\sigma$.  This means that
we will need some measure of the
response; i.e. some choice of a function
$\phi$ in \ref{response}.  In order
to weight all pointers equally, we choose
$\phi(\lambda,\tau)=1$.

We therefore associate the ``detection"
of a particle in $R$ with the projection
$\Pi_>$ onto the part of the
spectrum of $\int_R d\lambda d\tau \
A_{\lambda,\tau}$ greater
than or equal to $n\sigma$ and we
associate the projection
$\Pi_< = \openone - \Pi_>$
with the lack of detection of a particle in
region $R$.  We then define the ``probabilities"
for the particle to be
(or not to be) detected in $R$ by an ideal measurement
to be the limit of the
norms of these projections of $|\psi\rangle$ in which
$n\rightarrow
\infty$ but $n\sigma \rightarrow 0$.  Equivalently,
we could define
the ``probabilities" for the alternatives through
the corresponding
limit of the decoherence functional
\begin{equation}
\label{decoB}
{\cal D}^B_{\alpha,\alpha'} = \langle \psi | \Pi_{\alpha}
\Pi_{\alpha'}
|\psi\rangle
\end{equation}
for $\alpha \in \{>,<\}$ although decoherence is trivial even
before the limits are taken
since $\Pi_>$ and $\Pi_<$ are commuting projection operators.

In order to compute these ``probabilities," we note that the
expansion of
$\Pi_>|\psi\rangle$ (or $\Pi_<|\psi\rangle$)
in terms of the basis $|x,A;T_2\rangle$ of
simultaneous eigenvectors of $\hat{X}_g(T_2)$ and
$\hat{A}_{\lambda,\tau,g}
(T_2)$ is given by the coefficients
\begin{eqnarray}
\langle x_2,A_2;T_2|\Pi_{\alpha}|\psi\rangle &=&
\theta_{\alpha}\bigl[
\int_R d\lambda d\tau \ A_{\lambda,\tau,2} - n
\sigma \bigr] \
\int \prod_t dx \ \int \prod_{\lambda,\tau,t}
(dA_{\lambda,\tau}
d\pi_{\lambda,\tau}) e^{iS_B} \cr &\times&
{1 \over N} \exp{(-\int_R d\lambda d\tau
A_1^2/2\sigma^2)} \phi_p(x_1)
\end{eqnarray}
where $\phi_p(x) = \langle x;T_1|\phi_D\rangle$,
$\theta_>$ and $\theta_<$ are step functions with
support on the
positive and negative axes respectively, and the
sum is over all
paths $x(t)$ from $(x_1,T_1)$ to $(x_2,T_2)$
and all pointer configurations
and momenta subject to the boundary conditions $A_{\lambda,\tau}
(T_1) = A_{\lambda,\tau,1}$ and $A_{\lambda,\tau}
(T_2) = A_{\lambda,\tau,2}$.  Because of the trivial evolution that
we have assumed for the pointer degrees of freedom, the
integrations over the fields $A$ and $\pi$
at intermediate times are particularly simple and their only effect
is to require that $A_{\lambda,\tau}(T_2) - A_{\lambda,\tau}(T_1)
= g\int_{T_1}^{T_2} dt \ {\cal P}_{\lambda,\tau}
= gI_{\lambda,\tau}[x(t)]$.
It follows that we can calculate the
``probabilities" ${\cal D}_{>>}$ and
${\cal D}_{<<}$ from a path integral expression which,
after a shift of the integration variables $A_{\lambda,\tau}$
by the amount $I_{\lambda,\tau}[x(t)]$, takes the form
\begin{eqnarray}
\label{integral}
{\cal D}_{<<} &=& \int \prod_t (dx_1dx_2)
\ e^{i(S_{free}[x_1]-S_{free}[x_2])}
|\phi_p(x_1)|^2 \cr &\times& {1 \over N^2}
\int \prod_{\lambda,\tau} dA_{\lambda,\tau}
\ \exp(\int_R d\lambda  d\tau \  A_{\lambda,\tau}^2/\sigma^2)
\ \theta_<\bigl[\int_R d\lambda d\tau \ (A_{\lambda,\tau}
+gI_{\lambda,\tau})
- n \sigma\bigr]
\end{eqnarray}
Note that for a given path $x(t)$ that spends a total
time $T$ in the
region $R$, the integral over the pointer fields
(specifically, the
expression on the second line) gives just the
gaussian measure
(with width $\sigma$) of the set $S_{gT-n\sigma}$
of all configurations
such that $\int_R d\lambda d\tau \ A_{\lambda,\tau} + gT
- n \sigma < 0$, which we will denote by
$\mu_{\sigma}(S_{gT-n\sigma})$.

Now, consider a path such that $T>0$.  Note that
for small $n\sigma$ we have
$\mu_{\sigma}(S_{gT-n\sigma}) < \mu_{\sigma}(S_{gT/2})$
and that
as $\sigma \rightarrow 0$, the measure $\mu_{\sigma}$
is concentrated
on configurations near $A_{\lambda,\tau} = 0$.  It follows that
$\mu_{\sigma}(S_{gT-n\sigma}) \rightarrow 0$ in
the ideal measurement
limit where $n \rightarrow \infty$
but $n \sigma \rightarrow 0$
and that paths which enter $R$ do not contribute to
\ref{integral} in this limit.  Note that
we must, however, keep the coupling constant $g$ fixed
or send it to zero
more slowly than $\sigma$ in order for this conclusion to be
reached.  In this way, the limit of small coupling does not commute
with the limit of ideal measurement.

However, for a path with $T=0$ the integral over pointer
configurations gives the value
$\mu_{\sigma}(S_{-n\sigma})$ which approaches $1$ in the
limit $n \rightarrow \infty$, even when $n\sigma \rightarrow 0$.
Thus we find that in the
``ideal measurement limit" the ``probability for the particle to
avoid the region $R$," is given by the expression
\begin{equation}
\label{comp}
{\cal D}_{<<} = \int_{paths \ \cap \ R = \emptyset}
 \prod_t (dx_1dx_2) \ e^{i(S_{free}[x_1]-S_{free}[x_2])}
|\phi_p(x_1)|^2
\end{equation}
which can be related to the analysis of \cite{Hartle,first,YT}.

\subsection{Probabilities by Partitions of Paths}

The above expression was presented in refs. \cite{Hartle,first,YT}
as part of a proposal for the definition of probabilities for the
particle to enter or
avoid the region $R$ through partitions of paths in a path integral.
Specifically, \cite{Hartle,first,YT} associate the operator
 ${\cal C}_R$ (${\cal C}_{\overline{R}}$) in the Hilbert space
${\cal H}_{F:I}$ of an isolated free particle
with the alternative ``particle does (not) enter R," where

\begin{equation} \label{CRbar} \langle x_2;t_2|{\cal
C}_{\overline{R}}|x_1;t_1\rangle = \int_{paths \ \cap R =
\emptyset} e^{iS_{free}}
\end{equation}
and ${\cal C}_R \equiv \openone - {\cal
C}_{\overline{R}}$.
Here, the sum is over all
paths that begin at $(x_1,t_1)$ and end at  $(x_2,t_2)$
without passing through the region $R$, $S_{free}$ is
the action for the nonrelativistic free particle,
$|x;t\rangle$ is an eigenstate of the
time-dependent position operator $\widehat{X(t)}$ with
eigenvalue $x$,
and $t_1$ and $t_2$
are any times respectively to the past and future of $R$.

Probabilities are then defined through the decoherence functional:

\begin{equation}
\label{HYT}
{\cal D}^{HYT}_{\alpha, \alpha'} = \langle \phi | {\cal
C}_{\alpha}^{\dagger} {\cal C}_{\alpha} |\phi \rangle
\end{equation}
for $\alpha \in \{ R, \overline{R} \}$.
Although ${\cal C}_R
\equiv \openone - {\cal C}_{\overline{R}}$, decoherence is
not immediate since neither ${\cal C}_R$ nor ${\cal
C}_{\overline{R}}$ is a projection operator.
Since we will mention the
decoherence functional \ref{HYT} several times,
it will be convenient to refer
to it as the ``HYT
(Hartle-Yamada-Takagi) decoherence functional."

We note that ${\cal D}^B_{<<} = {\cal D}_{\overline{R},\overline{R}}
^{HYT}$, so that
${\cal D}_{>>}$ must also agree with ${\cal D}_{RR}^{HYT}$ whenever
the alternatives decohere for both decoherence functionals.
Since the integral \ref{CRbar} has been
calculated before, we quote the
result of \cite{Hartle,first} that
any ``probability"
between zero and one may be found,
depending on the state $|\psi\rangle$.  We then note that
this result is
quite different from what
we found in section \ref{I} but do not concern
ourselves further with calculation of expression
\ref{comp}.

As mentioned above,
$\Pi_>$ and $\Pi_<$ always decohere since they are
commuting projection operators,
although the same is not true of
${\cal C}_R$ and ${\cal C}_{\overline{R}}$.
As a result, ${\cal D}^B$
and ${\cal D}^{HYT}$ are not identical.
This difference is investigated in
the following subsection
and may be summarized by saying that the
decohering effect of device B is not
included in the HYT decoherence
functional.

\subsection{The disturbing nature of device B}
\label{disturbs}

At first glance it may be tempting to say that, since
\ref{decoB} is a decoherence functional for the device while the
HYT result concerns the free particle, we should not
be surprised that the two decoherence functionals ${\cal D}^B$
and ${\cal D}^{HYT}$ do not agree.  However, this
statement is not entirely satisfactory since, in the ideal
measurement
limit, we {\em do} find exact agreement for the
corresponding decoherence
functionals in a von Neumann measurement
(see appendix B for an
illustration involving device A).  The explanation lies in the
fact that, as pointed out in \cite{first},
we in general expect a measurement that takes place
over an extended
time to disturb the system being measured.  We now show
in detail how this
occurs in model B and how it accounts for the discrepancy between
${\cal D}^B$ and ${\cal D}^{HYT}$.

To do so, we first note that model
$A$ associates with $\Pi_>$ the projection
operator $\Pi_{I_g^-\hat{\chi}_0
(I_g^-)^{-1} \leq 0}$
and with $\Pi_<$ the projection operator $\Pi_{I_g^-\hat{\chi}_0
(I_g^-)^{-1} > 0}$
in the sense that
\begin{eqnarray}
{\cal D}^A_{\alpha,\alpha'} \equiv
\lim_{m \rightarrow i}
\langle \psi| \Pi_{\alpha} \Pi_{\alpha'}|\psi \rangle &=&
\lim_{{(n\sigma) \rightarrow 0} \atop {n \rightarrow \infty}}
\langle \psi| \int_{\chi'\in {\cal R}} \theta_{\alpha}
(\chi'-n\sigma) \Pi_{I_g^-\hat{\chi}_0
(I_g^-)^{-1}}
d\mu_{\hat{\chi}_f}(\chi') \cr &\times&
\int_{\chi'' \in {\cal R}} \theta_{\alpha'}
(\chi''-n\sigma) \Pi_{I_g^-\hat{\chi}_0
(I_g^-)^{-1}} d\mu_{\hat{\chi}_f}(\chi'') |\psi\rangle
\end{eqnarray}
where the notation $\lim_{m\rightarrow i}$ refers
to the limit in which
the measurement becomes ideal, the projection operators
correspond to the operator $I_g^-\hat{\chi}_0(I_g^-)^{-1}$,
the spectral measure corresponds to the operator
$\hat{\chi}_f$, and
$\alpha,\alpha' \in \{>,<\}$.
Since $\Pi_{\hat{\chi}_0 = \chi'} = \Pi_{\hat{\chi}_f =\chi'}
\otimes \openone_{0,D}$
and  $\openone -
\Pi_{\hat{\chi}_0 = \chi'} = (\openone - \Pi_{\hat{\chi}_f =\chi'})
\otimes \openone_{0,D}$
in terms  of the projection operator $\Pi_{\hat{\chi}_f =\chi'}$
onto eigenvalues of $\hat{\chi}_f$,
the naive decoherence functional \ref{naive}
defined by the state $|\phi_p\rangle$
and the projections $\Pi_{\hat{\chi}_f =\chi'}$
and $\openone - \Pi_{\hat{\chi}_f =\chi'}$ agrees with ${\cal D}^A$.

We can now contrast this line of reasoning
with the corresponding deductions for device B.
Again, we found that the desired decoherence functional could be
computed using operators associated only with the particle and not
with the device.  Specifically, in the ideal measurement limit we
found that
\begin{equation}
\lim_{m \rightarrow i}
\langle \psi| \Pi_{\alpha} \Pi_{\alpha'}|\psi \rangle =
\langle \psi| {\cal C}_{g,\alpha}
{\cal C}_{g,\alpha'}|\psi \rangle
\end{equation}
for $\alpha \in \{>,<\}$ where ${\cal C}_{g,<}$ is defined by its
matrix elements:
\begin{equation}
\label{factC}
\langle x_2,A_2;T_2|{\cal C}_{g,<}|x_1,A_1;T_1\rangle
= \delta(A_2-A_1) \int_{paths \ \cap \ R = \emptyset} \prod_t dx
e^{iS_{free}}
\end{equation}
and the sum is over all paths that begin at $(x_1,T_1)$ and
end at $(x_2,T_2)$.  The complimentary operator ${\cal C}_{g,>}$
is then defined as $\openone - {\cal C}_{g,<}$.
Since
$|x_1,A_1;t\rangle = |x_1;t\rangle \otimes |A_1;t\rangle$
with respect to the {\em t-dependent} factorization ${\cal H}_g
= {\cal H}_{g,p:t}\otimes {\cal H}_{g,D:t}$,
Eq. \ref{factC} implies
not that the operator ${\cal C}_>$ is
of the form ${\cal C}_f \otimes \openone_{0,D}$ for
some operator ${\cal C}_f$ in ${\cal H}_{g,p:T_1}$, but
that ${\cal C}_{g,<}$ is of the form:
\begin{equation}
{\cal C}_{g,<} = {\cal C}_{g,p,<} \otimes I_D
\end{equation}
where ${\cal C}_{g,p,<}$ maps ${\cal H}_{g,p;T_1}$
to ${\cal H}_{g,p;T_2}$
and $I_D$ is the isomorphism $I_D|A;T_1\rangle = |A;T_2\rangle$
from ${\cal H}_{g,D:T_1}$ to ${\cal H}_{g,D:T_2}$.

The relationship to the operators $\hat{\cal C}_{\overline{R}}$ and
$\hat{\cal C}_R$ that define the HYT decoherence functional can be
made clear through the introduction of two more isomorphisms
that relate the free particle Hilbert space ${\cal H}_{F}$ on which
$\hat{\cal C}_{\overline{R}}$ and
$\hat{\cal C}_R$ are defined to ${\cal H}_{g,p;T_1}$ and
${\cal H}_{g,p;T_2}$.  The isomorphisms
$I_{g,t}:{\cal H}_{F}\rightarrow{\cal H}_{g,p;t}$
for $t\in{T_1,T_2}$ are defined by
$I_{g,t}|x;t\rangle = |x;t\rangle$
so that ${\cal C}_{g,p,<} = I_{g,T_2} \hat{\cal C}_{\overline{R}}
I_{g,T_1}^{-1}$ and $\openone  -{\cal C}_{g,p,<} = I_{g,T_2}
(\openone - \hat{\cal C}_{\overline{R}})
I_{g,T_1}^{-1}$.  The subtle point is that
\begin{equation}
\label{neg}
{\cal C}_{g,>} = \openone - {\cal C}_{g,<} \neq   I_{g,T_2}
(\openone - \hat{\cal C}_{\overline{R}}) I_{g,T_1}^{-1} \otimes I_D
\end{equation}
That this is so can be seen by adding the right hand side above to
${\cal C}_{g,<}$:
\begin{equation}
I_{g,T_2} (\openone -
\hat{\cal C}_{\overline{R}})  I_{g,T_1}^{-1}\otimes I_D
+ {\cal C}_{g,<}
= I_{g,T_2}I_{g,T_1}^{-1} \otimes I_D
\end{equation}
but $ I_{T_2}I_{T_1}^{-1} \otimes I_D$ is the isomorphism between
${\cal H}_{p;T_1}\otimes{\cal H}_{D;T_1}$ and
${\cal H}_{p;T_2}\otimes{\cal H}_{D;T_2}$
induced by the corresponding
factorizations of the Hilbert space ${\cal H}_0$
of the {\em uncoupled} system.  This isomorphism
therefore differs from the
identity operator in any ${\cal H}_g$ for which
the coupling constant is nonzero.

How does this all relate to our characterization of device B as
a ``disturbing" apparatus?  We note that the operators picked
out by device A were of
the form $\Pi_{I_g^-\hat{\chi}_0 (I_g^-)^{-1}}
= (I^-_g)^{-1}
\Pi_{\hat{\chi}_0 = \chi'} I^-_g$ and hence were ``undisturbed."
Suppose the
same were true of ${\cal C}_{g,<}$, i.e. that
${\cal C}_{g,<} = I^-_g
{\cal C}_{0,<} (I^-_g)^{-1}$.  In this case, since
expression \ref{factC} holds for all $g$, it holds in particular
for the uncoupled system $g = 0$.  Since, for the
uncoupled system \ref{neg} is an equality, it follows that
\begin{equation}
{\cal C}_{g,>} =   I^-_g \ I_{0,T_2} (\openone -
\hat{\cal C}_{\overline{R}}) I_{0,T_1}^{-1} \otimes I_D (I^-_g)^{-1}
\end{equation}
and therefore that the decoherence functionals
${\cal D}^B$ and ${\cal D}^{HYT}$ are identical.  Since this is
exactly the
line of reasoning used above to relate ${\cal D}^A$ to the
``naive" result, we
attribute the discrepancy between ${\cal D}^B$ and ${\cal D}^{HYT}$
to the fact that this reasoning is not applicable here.  That is,
we attribute the difference to the fact that the ``response"
of device
B is to an operator ${\cal C}_{g,<}$ which differs from its
uncoupled version ${\cal C}_{0,<}$ so that we may
refer to it as ``disturbed" by device B.

\section{Discussion}
\label{Dis}

We have analyzed two ``devices" designed to detect the
presence of a particle in a spacetime region $R$.  These
two devices lead to quite different interpretations of
the question  ``What is the probability
that a quantum free particle will be detected in a spacetime region
$R$" as is illustrated by the different calculations
performed and different results obtained in sections
\ref{I} and \ref{II}.  We therefore conclude that this question
is not well-defined without reference to the kind of
detector that will be used and that different prescriptions for the
calculation of probabilities are appropriate to different physical
situations.  In particular, since ${\cal D}^{HYT}$ is associated only
with device B and not device A, this study indicates the type of
situations to which probabilities defined by
partitions of configuration
space paths in a path integral are and are not appropriate, although
the sense in which ${\cal D}^{HYT}$ is associated with
device B is not quite the usual one.

Which of our models and therefore which calculation would
be relevant to an actual experimental setting will
depend on the experimental details, although Appendix A suggests
that the most likely answer is
``neither A nor B."  Which model is more relevant to
philosophical discussions of ``measurements in spacetime
regions" will be subject to the interpretations of the
philosopher.  We may, however, make the distinction that
device A performs a single von Neumann measurement while device B
does not and that device A responds to an
undisturbed value of $\chi$
while the response of B is related to the disturbed operators
${\cal C}_>$ and  ${\cal C}_<$.

Finally, we would like to point out that the model
discussed in \ref{II} is {\em our interpretation} of the
HYT decoherence functional and we have
given no proof that this interpretation is unique.  The
author would, however, be willing to conjecture
that the only interpretation
of ${\cal D}^{HYT}$ given by a study of detectors is
more or less the one that we have described,
provided that the various terms in
this statement can be more or less precisely defined.
In support of
this conjecture, note that \cite{Hartle} describes such
decoherence functionals as intuitively related to
an infinite product
of projection operators
and the calculation of such path integrals in
\cite{first} uses a skeletonized version of the product:
$\prod_t \widehat{\theta(x(t))}$.
Such a product is naturally related
to models similar to B in which a large number
of independent interactions
take place at successive instants of time.  Any further
clarification of this issue, either by a formalization of
the above conjecture and subsequent proof or by a description of
other measurement models that provide an
alternate interpretation of the
HYT decoherence functional
would be much appreciated.

\acknowledgements
This work was partially supported by NSF grant
PHY-9005790 and by research funds provided by Syracuse
University.  The author would like to express his thanks
to Guillermo
Mena Marug\' an and Jorma Louko
for helpful editorial comments and to Jim
Hartle for useful discussions.

\appendix

\section{Bubble Chambers}
\label{real}

Neither apparatus described in
sections \ref{I} and \ref{II}
bears much resemblance to common measuring
devices actually used to detect particles in spacetime
regions.   In this appendix also, we will not discuss an
{\em accurate} model of a real such device, nor do we
analyze the simplistic ``model" of a
bubble chamber that we do present in the
manner of sections \ref{I} and \ref{II}.  We describe
this model only to show that neither device (A or B) models an
apparatus resembling a bubble chamber and
choose the bubble chamber for
its historical value as the following discussion applies
equally well (and equally poorly) to photographic film,
scintillation counters, etc. and

Each localized group of molecules in a bubble chamber
constitutes a degree of freedom: a group either nucleates a
bubble or it doesn't.  Let us assume that the chamber is
``cold" and that these molecules do not move significantly
during the experiment  so that we may associate one degree
of freedom with each point of the space inside the
chamber.  Each of these degrees of freedom is sensitive to
the amount of time that the particle spends in its
vicinity, since the longer the particle stays nearby, the
more likely it is that a bubble will form.  In order to
relate bubble chambers to the ``devices" described above,
we might construct a model which has one ``pointer"
variable $A_{\lambda}$ for each point $\lambda$ in the
chamber, such that each of these pointers interacts with
the particle through a von Neumann like term when the
particle occupies its position between the times
$T_1$ and $T_2$.  We might choose, for example, the action:
\begin{equation} S_{tot} = S_{free} +
\int_{-\infty}^{\infty} dt \int_{\lambda \in \ chamber}
d\lambda \ \pi_{\lambda} \delta(x - \lambda) +
\int_{-\infty}^{\infty} dt \int_{\lambda \in \  chamber}
d\lambda dt \ \pi_{\lambda}\dot{A}_{\lambda} \end{equation}
Such a  model couples pointers to the
integrals ${\cal I}_{\lambda}
= \int_{-\infty}^{\infty}dt
\int_{-\infty}^{\infty} d\tau \
{\cal P}_{\lambda,\tau}(t)$ over times
$\tau\in(T_1,T_2)$ instead of to the integrals
${\chi} = \int_{-\infty}^{\infty} dt
\int_R d\lambda d\tau {\cal P}_{\lambda,\tau}(t)$
over all of $R$ as in model
A or to the individual $\int_{-\infty}^{\infty} dt
\ {\cal P}_{\lambda,\tau}$
as in model B.  Thus, we see that
neither of the
devices described in sections \ref{I} and \ref{II}
models such a device.

We should comment  once again that even the model
just presented is
far from an accurate description of a real bubble chamber.
An important feature of the real device is its
discrete nature: the degree of freedom at a given point
either ``triggers" (nucleates a bubble) or does not so
that the bubble nucleation proceeds probablistically.
For a
classic description of bubble chambers which does capture
this discrete feature,  see \cite{Mott}.

\section{The ideal measurement limit}
\label{limitA}

In this appendix, we briefly review how the standard ``probability"
distribution for measurements is
reached in the ideal measurement limit.  Specifically, we derive
the result \ref{distributions}\footnote{Strictly speaking we derive
the result \ref{distributions} only near $A'=0$,
though the general result
follows in the same way.  In addition,
we note that Eq. \ref{distributions}
is only needed near $A'=0$ for the
discussion of section \ref{I}.}.  We
begin by assuming that state of the
pointer at time $T_1$
is $|g_{\sigma}(0);g,T_1\rangle$ peaked
around $A_{\lambda,\tau}'=0$ where
\begin{equation}
\label{gaussA}
|g_{\sigma}(A');g,T_1\rangle =
\int_{-\infty}^{\infty}dA {{e^{-(A-A')^2/2
\sigma^2}} \over {\pi^{1/4}\sqrt{\sigma}}} |A;g,T_1\rangle
\end{equation}
is a normalized state in ${\cal H}_{g,D:T_1}$, though the precise
form \ref{gaussA} will not be essential to our discussion.
Since the initial state of the pointer has a finite width $\sigma$,
we would like the pointer to move at least this far to the right
before we can confidently say that a particle has been detected.
Let us therefore define
``detection" of the particle as the presence of the pointer at least
$n\sigma$ to the right of the origin at time $T_2$ so that we will be
concerned with the projections $\hat{\Pi}_{A(T_2)<n \sigma}$
and $\hat{\Pi}_{A(T_2)>n \sigma}$
onto the appropriate part of the spectrum of $\int_R d\lambda d\tau \
A_{\lambda,\tau}(T_2)$.
Decoherence is then immediate since
these projections commute.  Note that classically we might say that
our ``confidence that
the pointer has responded to the presence of a particle"
is determined
by $n$, while the ''absolute inaccuracy" in
the measurement is given by
$n\sigma$.

Now, the pointer values before and after
the experiment are related by:
\begin{equation}
\label{ev}
\hat{A}_g(T_2) = \hat{A}_g(T_1) + gI^-_g\hat{\chi}_0 (I^-_g)^{-1}
\end{equation}
so that, if the state of our system is
\begin{equation}
|\psi\rangle = |\phi\rangle \otimes |g_{\sigma}(0);g,T_1\rangle
\end{equation}
for some normalized $|\phi_p\rangle \in {\cal H}_{f:T_1}$,
it is also of the form:
\begin{equation}
|\psi\rangle = \int d\mu_{\hat{\chi}_f}(\chi_1) \ \phi(\chi_1)
|\chi_1\rangle \otimes |g_{\sigma}
(g\chi_1);g,T_2\rangle
\end{equation}
in terms of a factorization defined by the commuting
operators $I_g^- \hat{\chi}_0 (I_g^-)^{-1}$ and $\hat{A}_g(T_2)$.
Here, $|g_{\sigma}(\chi_1);g,T_2\rangle$ is a normalized
state defined in analogy with $|g_{\sigma}(\chi);g,T_1\rangle$
of Eq. \ref{gaussA} and $|\chi_1\rangle$ is the eigenstate of
${I}_g^-\hat{\chi}_0(I_g^-)^{-1}$ with eigenvalue $\chi_1$.
The measure
$d\mu_{\chi_f}(\chi_1)$ is the spectral measure
for the operator $\hat{\chi}_f$
and the function $\phi(\chi_1)$ is given by
$\phi(\chi_1) = \langle \chi_1;T_1 | \phi \rangle$, where
$|\chi_1;T_1\rangle$ is the eigenvector of $\hat{\chi}_f$
with eigenvalue
$\chi_1$ in ${\cal H}_{g,p:T_1}$.
The projection that corresponds to the ``avoids $R$" alternative is
then:
\begin{equation}
\hat{\Pi}_{A(T_2)<n\sigma} |\psi\rangle =
\int d\mu_{\hat{\chi}_f}(\chi_1) \ \phi(\chi_1)
|\chi_1\rangle \otimes |h_{\sigma}
(g\chi_1);1,T_2\rangle
\end{equation}
where
\begin{equation}
|h_{\sigma}(g\chi);T_2\rangle
= \int_{-\infty}^{n\sigma} dA\ {{e^{-(A-g\chi)^2/2
\sigma^2}} \over {\sqrt{2\pi} \lambda}} |A;T_2\rangle
\end{equation}
and the ``probability" for detecting the particle in region $R$ is
defined by
\begin{equation}
{\cal P}_{R,\sigma} = \langle \psi |
\hat{\Pi}_{A(T_2)<n\sigma}|\psi\rangle
= \int d\mu_{\hat{\chi}_f}(\chi) \ \int_{-\infty}^{n\sigma} dA
{{e^{-(A-g\chi)^2/
n\sigma^2}} \over {\sqrt{2\pi} \sigma}}  |\phi(\chi)|^2
\end {equation}
Note that for any $n>0$
this goes to zero in the limit $\sigma \rightarrow 0$ since
$\hat{\chi}_f$ has no normalizable eigenvectors with eigenvalue
$\chi \leq 0$.  Note also that we keep the coupling constant finite
while taking this limit or at least, if the limit of small
coupling is to be
taken, we must send $g$ to zero mote slowly than more
slowly than $\sigma$.  In this way, we see that the limits of small
coupling and ideal measurement do not commute.

Since $\sigma \rightarrow 0$ is the limit in which the measurements
become more and more accurate, we say that the probability
is {\em one}
for a perfect
detector to find the particle in region $R$ with
any confidence level
$n$.  We may in fact take the limit $n \rightarrow \infty$
simultaneously with the limit $\sigma \rightarrow 0$ so long as
the ``absolute inaccuracy" $n\sigma$ becomes zero
in this limit as well.
We therefore refer to
the limit $n \rightarrow 0$, $n\sigma \rightarrow
0$ as the ``ideal measurement limit."

\end{document}